\documentclass[prd,aps,twocolumn,showpacs,preprintnumbers,amsmath,amssymb]{revtex4}
\usepackage[dvips]{graphicx}
\usepackage{epsf}
\usepackage{amsmath}
\usepackage{amssymb}

\usepackage{graphicx}
\usepackage{dcolumn}
\usepackage{bm}
\def\be{\begin{equation}}
\def\ee{\end{equation}}
\def\bea{\begin{eqnarray}}
\def\eea{\end{eqnarray}}

\newcommand{\comment}[1]{}

\newcommand{\tphi}{\tilde{\phi}}

\begin{document}


\date{\today}

\title{The Trans-Planckian Problem for Inflationary Cosmology Revisited}

\author{Robert Brandenberger$^{1,2,3}$ and Xinmin Zhang$^{2,3}$}

\affiliation{1) Department of Physics, McGill University,
Montr\'eal, QC, H3A 2T8, Canada}

\affiliation{2) Institute of High Energy Physics, Chinese Academy of Sciences, P.O. Box 918-4,
Beijing 100049, P.R. China}

\affiliation{3) Theoretical Physics Center for Science Facilities (TPCSF), Chinese Academy of
Sciences, P.R. China}


\begin{abstract}

We consider an inflationary universe model in which the phase of accelerated
expansion was preceded by a non-singular bounce and a period of
contraction which involves a phase of deceleration. We follow 
fluctuations which exit the Hubble radius in the radiation-dominated contracting
phase as quantum vacuum fluctuations, re-enter the Hubble radius in the
deflationary period and re-cross during the phase of inflationary expansion.
Evolving the fluctuations using the general relativistic linear perturbation
equations, we find that they exit the Hubble radius during inflation not with
a scale-invariant spectrum, but with a highly red spectrum with index $n_s = -3$.
We also show that the back-reaction of fluctuations limits the time interval of deflation.
Our toy model demonstrates the importance for inflationary cosmology
both of the trans-Planckian problem for cosmological perturbations and of back-reaction effects . 
Firstly, without understanding both Planck-scale physics and the phase which preceded inflation, 
it is a non-trivial assumption to take the perturbations to be in their local vacuum state when they
exit the Hubble radius at late times. Secondly, the back-reaction effects of
fluctuations can influence the background in an important way.

\end{abstract}

\maketitle

\newcommand{\eq}[2]{\begin{equation}\label{#1}{#2}\end{equation}}

\section{Introduction}

As was pointed out in \cite{RHBrev0} and discussed in detail in
\cite{Jerome} (see also \cite{Niemeyer}), the usual computation of
the spectrum of fluctuations in inflationary cosmology (see
\cite{Mukh} for the original reference, \cite{MFB} for a
comprehensive review, and \cite{RHBrev1} for a more
condensed overview) suffers from a serious conceptual
problem: provided that the duration $\Delta t$ of inflation
exceeds $70 H^{-1}$, where $H$ is the Hubble constant
during inflation, then the physical wavelengths of all
scales which are being probed today through observations
were smaller than the Planck length at the beginning of the
period of inflation \footnote{The numerical coefficient $70$
depends mildly on the scale of inflation: we have chosen
the scale to be that of Grand Unification, the scale which is
required for simple inflaton potentials.}. The computation of
the spectrum of cosmological perturbations is based on
General Relativity and classical scalar field theory, both
of which are clearly not valid (and not even good approximations)
on length scales smaller than the Planck length. Thus, in
inflationary cosmology the fluctuations emerge from the
``trans-Planckian region of ignorance" (see Figure 1).

\begin{figure}[htbp]
\includegraphics[scale=0.5]{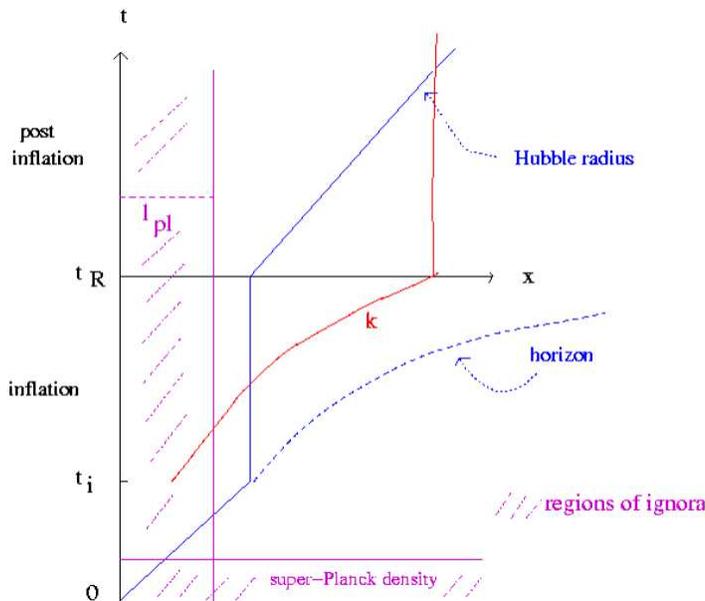}
\caption{Illustration of the trans-Planckian problem for fluctuations
in inflationary cosmology. If the period of inflation is sufficiently long,
then at the beginning of the inflationary phase the physical wavelength
of all scales $k$ which are being observed today is smaller than
the Planck length $l_{pl}$. In the plot, the vertical axis is time, and the
horizontal axis denotes physical distance. The period of inflation lasts
from $t_i$ to the time of reheating $t_R$. The dashed regions 
indicate zones where one cannot trust the effective equations of
motion derived from General Relativity. These break down on length
scales smaller than the Planck length, and at energy densities which
are sufficiently close to the Planck density $m_{pl}^4$. }
\end{figure}

One might have hoped that the predictions of inflation would be
robust to how one models the physics of the trans-Planckian
scales. However, it is easy to construct toy models of
trans-Planckian physics in which large changes to the
usual predictions arise \cite{Jerome} \footnote{By large
we mean order unity.}. One can, however, impose
initial state criteria for fluctuations in inflationary models
which ensure that there are only minor modifications to
the usual predictions of inflation, e.g. by imposing a
time-like ``new physics hypersurface" which lies outside
of the trans-Planckian zone of ignorance, and demanding
that fluctuations originate in a state which minimizes the
energy at that time \cite{NPH}.

A further conceptual problem which scalar field-driven models
of inflation do not solve is the singularity problem. As shown
in \cite{Borde}, a past singularity is unavoidable in scalar field-driven
models of inflation.

Recently, there have been a lot of attempts to resolve the
cosmological singularity by means of a cosmological bounce. Such a
bounce can be achieved by considering a particularly chosen higher
derivative gravity action which is free of ghosts \cite{BMS,BBMS},
and it can be obtained by introducing quintom matter
\cite{quintom} \footnote{A particular realization of the quintom
bounce can be obtained by considering the scalar field sector of the
Lee-Wick model \cite{LW}.}. Ghost condensation with additional
higher derivative terms may also lead to a non-singular bounce (see
e.g. \cite{ghost}). Bounces can also be obtained by making use
of the spatial curvature term in the context of Einstein gravity
\cite{Starob78}. The minimal length principle of string theory
also points to a resolution of the singularity problem \cite{BV},
possibly involving a bouncing cosmology. Finally, both brane
world models \cite{Varun} and loop quantum
cosmology  (see \cite{LQC} for a recent review)
can give rise to non-singular bounces (for a recent
general review on bouncing cosmologies see \cite{Novello}).

The first point we wish to make in this Note is that non-singular
bouncing cosmologies with a post-bounce inflationary phase
provide a clean framework to
discuss the trans-Planckian problem of inflationary cosmology:
the fluctuations can be defined at early times during the
contracting phase when their physical wavelength is
in the far infrared. The fluctuations then evolve in a
non-singular way through the bounce into the period of
inflation (for previous studies of the impact of pre-bounce
evolution on the spectrum of cosmological perturbations in
inflationary cosmology see \cite{Joras,Xinmin}). 
In a model which includes a period of deflation before
the bounce, we demonstrate that, at least for a range of modes,
the fluctuations will not be in a Bunch-Davies-like
vacuum state on sub-Hubble scales during the phase
of inflationary expansion. Thus, the trans-Planckian problem
for inflationary fluctuations is revealed to be a very serious
one.

The second point which we make in this Note is that a
pre-bounce phase of exponential contraction is unstable
to the back-reaction effects of fluctuations which enter the
Hubble radius during the contraction. Due to the fact that
fluctuations grow on super-Hubble scales, they rather
generically obtain a red spectrum, and lead to a back-reaction
energy-momentum tensor which dominates over the
vacuum contribution leading to deflation. This leads
to a constraint on the range of modes for which large
trans-Planckian effects are predicted 
(see e.g. \cite{Tanaka,Starob,Jerome3,Kolb} for work pointing out
that demanding that inflationary expansion is robust
puts constraints on the amplitude of trans-Planckian
effects). 

The outline of this note is as follows. In the following section we
review the trans-Planckian problem for inflationary cosmology
and give a brief overview of previous approaches. In Section 3
we present a particular bouncing cosmology with a phase of
inflationary expansion. We compute the spectrum of fluctuations
starting with a Bunch-Davies vacuum state in the early stages
of the contraction. We find that the squeezing of the fluctuations
on a super-Hubble scales during the period of contraction leads
to a spectrum of fluctuations which is not close to scale-invariant.

\section{The Trans-Planckian Problem for Cosmological Perturbations}

The initial discussions of the trans-Planckian problem for fluctuations
were based on considering non-trivial dispersion relations for modes
at frequencies larger than the Planck length \cite{Jerome}. It was
found that suitable choices of dispersion relations can lead to large
trans-Planckian effects, including order unity changes in the spectral
index of cosmological perturbations (see e.g. \cite{Jerome2}). What is
happening is that modes evolve non-adiabatically on length scales
smaller than the Planck length. The time interval which modes spend
at sub-Planck length scales depends on the wave-number, and
hence short wavelength modes are more excited than long wavelength
modes. The magnitude of the trans-Planckian effects is
constrained by demanding that the energy in short wavelength modes
does not prevent inflation \cite{Tanaka,Starob,Jerome3}. In this case,
interesting trans-Planckian effects with amplitude of the order
$H / m_{pl}$ for scalar metric fluctuations can be found.

If one assumes that modes exit the Hubble radius in the local
vacuum state \cite{Kaloper1}, then the trans-Planckian effects
are very small (namely of order $(H / m_{pl})^2$. However, the
whole point of the trans-Planckian problem for fluctuations is
that there is no reason to expect that modes should be in the local
vacuum state when they exit the Hubble radius.

In another approach to the trans-Planckian problem, a ``new
physics hypersurface" can be introduced corresponding to
the Planck length, and it can then be assumed that fluctuations
emerge in the local vacuum state \cite{NPH}. In this case, the
amplitude of the trans-Planckian effects is bounded from above
by $H/m_{pl}$.

Imposing initial conditions for all modes at a fixed time
breaks the translation symmetry which ensures the scale-invariance
of the spectrum. But such effects are red-shifted as the duration of
the inflationary phase is increased (see e.g. \cite{Cline}) if the
calculations are done in the context of the usual inflationary
effective field theory.

Other approaches to the trans-Planckian problem include choosing
different initial state prescriptions (the so-called $\alpha$ vacua
\cite{alpha}), considering the effects of space-space
\cite{space} or space-time \cite{sptime} uncertainty relations
on the evolution of fluctuations, and studying the effects of
ultraviolet cutoffs \cite{Kempf}.  For a more comprehensive review
of different approaches to the trans-Planckian problem see
\cite{Jerome2}.

As we study in this Note, bouncing cosmologies with a phase
of inflationary expansion offer another setup to study the
trans-Planckian problem for inflationary fluctuations. This
issue was initially considered in \cite{Joras}, where it
was found that the trans-Planckian problem in models
with a modified dispersion relation persists in a bouncing
cosmology. In \cite{Xinmin}, a bouncing inflationary cosmology
was studied in which there is no phase of deflation before
the bounce. In this case, interesting signatures of the
pre-bounce physics in the spectrum of cosmological
fluctuations were found, but these effects were small in
amplitude.

In this Note, we consider a bouncing inflationary
cosmology which contains a period of deflation before the
bounce. In this case, a range of scales of interest crosses the Hubble radius
four times - first they exit the Hubble radius during the pre-deflation
phase, then they enter during the deflationary phase, they exit again during inflation,
to finally re-enter at late times. For these scales, we will find
very large trans-Planckian effects. In fact, it is the large effect
of these modes which limits the duration of the deflationary phase.

\section{Fluctuations in a Specific Bouncing Inflationary Cosmology}

\subsection{Background}

In this Note, we will be considering a bouncing inflationary cosmology
which contains a deflationary period before the bounce. 
The specific nature of the physics
providing the non-singular bounce will not be relevant for our
analysis. To be specific, we may assume that the bounce is provided
by the specific higher derivative gravity model of \cite{BMS,BBMS}.

We will model both the radiation phase and the inflationary phase
with a scalar field $\varphi$ with a potential
\be
V(\varphi) \, = \, \frac{\lambda}{4} \varphi^4 \, .
\ee
For field values $|\varphi| \ll m_{pl}$ (where $m_{pl}$ is the Planck
mass) the field is oscillating and the time-averaged equation of state
is that of radiation (see e.g. \cite{STB}), whereas for field values
$|\varphi| \gg m_{pl}$ the field is slowly rolling and hence leads to
an equation of state $w \sim -1$ which yields inflation \footnote{We
use the standard notation where the equation of state parameter $w$ is
the ratio $w = \frac{p}{\rho}$, with $p$ and $\rho$ being pressure
and energy density, respectively.}. We will first describe the
background dynamics in the absence of back-reaction of the
fluctuations. In the following section we will then discuss how the
back-reaction modifies the background dynamics.

We begin the evolution in a contracting radiation phase in which $\varphi$ is
oscillating with a small amplitude. Due to Hubble anti-friction during
the contracting phase, the field amplitude grows, eventually leading
to a transition to a deflationary phase with $w \sim -1$. This transition
takes place at a time which we denote by $- t_2$. Once the
energy density reaches the critical value where the physics that
determines the bounce sets in, the universe will undergo a
non-singular bounce, after which a period of inflationary expansion
will set in during which $\varphi$ is slowly rolling down its potential.
Once $|\varphi| \sim m_{pl}$, a smooth transition to a radiation phase
will set in. A space-time sketch of our scenario is given in Figure 2.

\begin{figure}[htbp]
\includegraphics[scale=0.3]{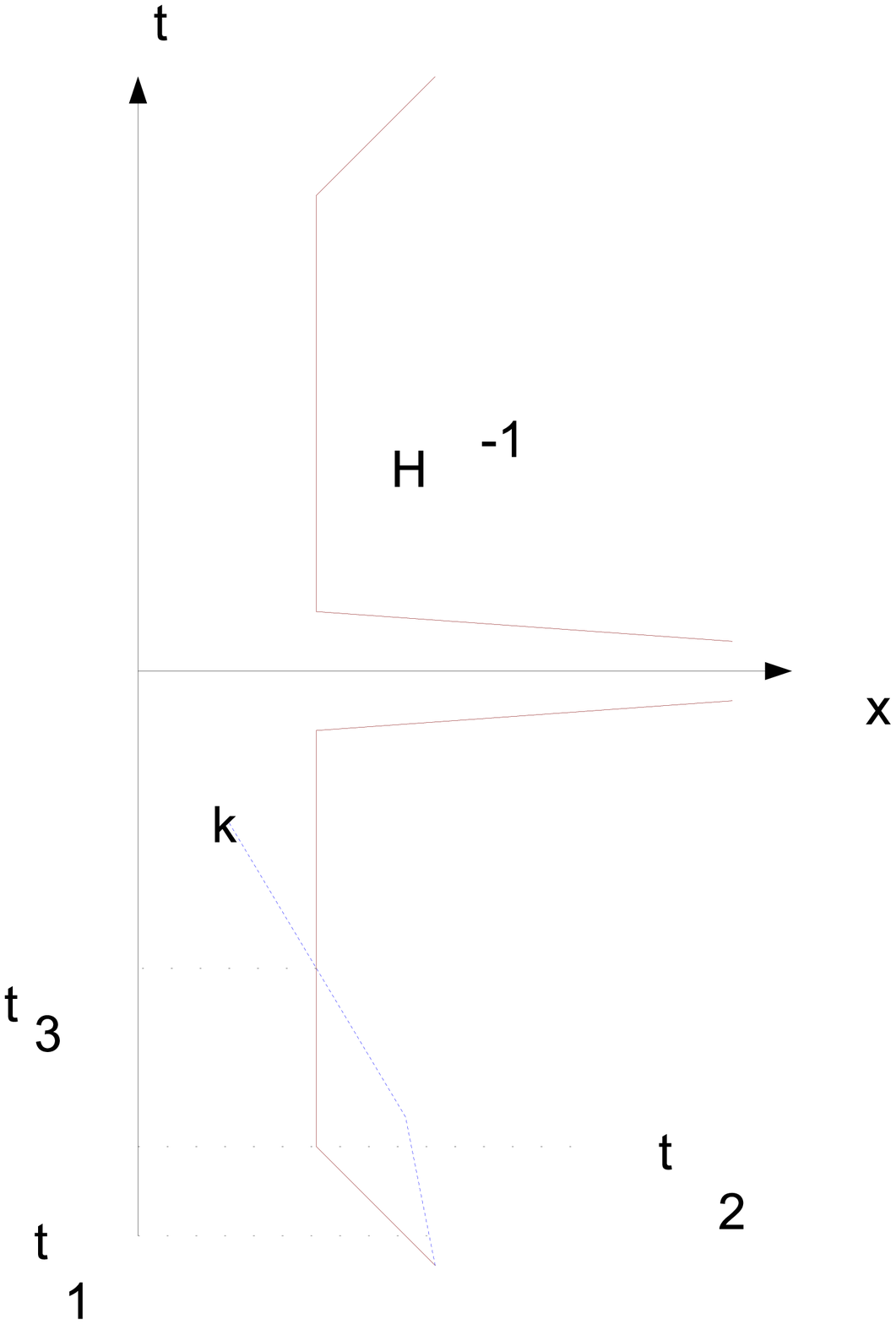}
\caption{Space-time sketch of our bouncing inflationary cosmology (in the
absence of the back-reaction of fluctuations). Time
is along the vertical axis, physical length along the horizontal axis. The
red curve denotes the Hubble radius $H^{-1}(t)$, the blue curve denotes
the physical wavelength corresponding to a fixed co-moving scale $k$.
This scale exits the Hubble radius during the phase of radiation-dominated
contraction at the time $-t_1(k)$ (the minus sign and the k-dependence are
not included in the figure), the transition to the deflationary phase occurs
at the time $-t_2$, and the scale re-enters the Hubble radius during the
deflationary phase at the time $- t_3(k)$ (once again, the minus sign and
the k-dependence are not included in the sketch).}
\end{figure}

As can be seen from Fig. 2, fluctuations originate on sub-Hubble
scales during the radiation-dominated contracting phase. They
exit the Hubble radius at time $- t_1(k)$ during this phase. We
are interested in scales which re-enter the Hubble radius during
the contracting deflationary phase (at a time which we denote by
$- t_3(k)$) and exit the Hubble radius
once again during the phase of inflationary expansion, before
finally re-entering the Hubble radius at late times (at the time
$t_1(k)$).

\subsection{Fluctuations: Formalism}

In the context of inflationary cosmology it has proven to be
convenient \cite{BST,BK,Lyth} to track the cosmological fluctuations
in terms of the variable $\zeta$, the curvature fluctuations in
co-moving coordinates. This variable is conserved at phase
transitions and is constant on super-Hubble scales in an
expanding universe. The variable $\zeta$ is related to the
metric fluctuation $\Phi({\bf{x}}, \eta)$ in longitudinal gauge (in which the
metric takes the form \footnote{We are neglecting anisotropic stress.}
\be
ds^2 \, = \, a^2(\eta) \bigl[ (1 + 2 \Phi) d\eta^2 - (1 - 2 \Phi) d{\bf{x}}^2 \bigr] \, ,
\ee
where $\eta$ is conformal time and the ${\bf{x}}$ denote co-moving spatial
coordinates) via
\be
\zeta \, = \, \frac{2}{3} \bigl( {\cal{H}} \Phi^{'} + \Phi \bigr) \frac{1}{1 + w} + \Phi \, ,
\ee
$\cal{H}$ denoting the Hubble expansion rate in conformal time, a prime
indicating the derivative with respect of $\eta$, and $w = p / \rho$ being
the equation of state parameter ($p$ and $\rho$ are pressure and
energy density, respectively).

The variable $\zeta$ is closely related to the variable $v$
\cite{MFB} in terms of which the action for cosmological
fluctuations has canonical kinetic term: 
\be
\zeta \, = \, \frac{v}{z} \,
\ee
where for hydrodynamical matter
\be
z(\eta) \, = \, \frac{1}{c_s \theta} \,
\ee
where $c_s$ is the speed of sound and
\be
\theta \, = \, \frac{\cal{H}}{a} \bigl[ \frac{2}{3} ({\cal{H}}^2 - {\cal{H}}^{'}) \bigr]^{-1/2} \, .
\ee
The equation of motion for the Fourier mode $v_k$ of $v$ is
\be
v_k^{''} + \bigl( k^2 - \frac{z^{''}}{z} \bigr) v_k \, = \, 0 \, .
\ee
This shows that on length scales larger than the Hubble radius,
where the $k^2$ term is negligible, $v$ is frozen in (i.e. it does
not oscillate), whereas on sub-Hubble scales $v_k$ is oscillating
with approximately constant amplitude.

The equation of motion for $v$ has a singularity at the bounce
point which is associated with the fact that at this point the
comoving gauge is not well-defined. This issue was discussed
for example in Refs. \cite{bounce}. However, the problems
are due to singularities in the factor $\frac{z^{''}}{z}$. Since
we are interested in scales which are sub-Hubble near the bounce,
this problem does not arise for us. Since the equation of motion
for $\zeta$ is easier to solve during the transition from radiation-domination
to deflation, we will follow $\zeta$.

\subsection{Fluctuations in the Contracting Radiation Phase}

On super-Hubble scales, the equation of motion for $v_k$ in a universe
which is contracting or expanding as a power $p$ of time, i.e.
\be
a(t) \, \sim \, t^p \, , \ee 
is given by
\be
v^{''}_k \, = \, \frac{p(2p -1)}{(p -1)^2} \eta^{-2} v_k \, ,
\ee
which has solutions
\be
v(\eta) \, \sim \, \eta^{\alpha}
\ee
with
\be \label{alpha}
\alpha \, = \, \frac{1}{2} \pm \nu \,\,\,\, , \,\,\,\, \nu \, = \, \frac{1}{2} \frac{1 - 3p}{1 - p} \, .
\ee

In the radiation phase we have $p = 1/2$ and hence $\nu = - 1/2$ which leads to
the two values for $\alpha$ which are $\alpha = 1$ and $\alpha = 0$. Hence,
\be
v_k(\eta) \, = \, c_1 \eta + c_2 \, ,
\ee
where $c_1$ and $c_2$ are constants, and hence (making use of
$a(\eta) \sim \eta$)
\be \label{rad}
\zeta_k(\eta) \, = \, c_1 + c_2 \eta^{-1} \, .
\ee
In a contracting universe, the second term dominates, and we conclude
that $\zeta$ is growing on super-Hubble scales.

\subsection{Fluctuations in the Contracting Deflationary Phase}

During deflation we have
\be
a(\eta) \, \sim \, \eta^{-1} \, ,
\ee
with $\eta$ tending to infinity as the bounce is approached. From (\ref{alpha})
we see that the two solutions of the equation for $v$ have values
\be
\alpha \, = \, 2 \,\,\,\, \rm{and} \,\,\,\, \alpha \, = \, -1 \, .
\ee
Hence,
\be
\zeta_k(\eta) \, = \, c_3 \eta^3 + c_4 \, ,
\ee
where $c_3$ and $c_4$ are constants. The first mode is the growing mode

Now we want to calculate the spectrum of fluctuations when they
enter the Hubble radius during the deflationary phase, starting with
vacuum initial conditions in the early contracting radiation phase, i.e. with
the spectrum
\be
\zeta(k, -t_1(k)) \, = \, \frac{v}{a}(k, -t_1(k)) \, \sim \, \frac{k^{-1/2}}{a(-t_1(k))}
\, \sim \, k^{1/2} \, ,
\ee
making use of the Hubble radius crossing condition in the radiation phase
\be
a(-t_1(k)) k^{-1} \, = \,  2 t_1(k)
\ee
which leads to
\be \label{scaling}
\eta(t_1(k)) \, \sim \, k^{-1} \, .
\ee

Next, we compute the spectrum at the end of the radiaton phase, making use
of (\ref{rad})
\be
\zeta(k, -t_2) \, = \, \frac{\eta(-t_1(k))}{\eta_2} \zeta(k, -t_1(k)) \, \sim \, k^{-1/2} \, ,
\ee
where we have made use again of (\ref{scaling}).

Finally, we compute the spectrum at Hubble radius re-entry in the deflationary
phase:
\be
\zeta(k, -t_3(k)) \,  = \, \bigl( \frac{\eta(-t_3(k))}{\eta(t_2)} \bigr)^3 \zeta(k, -t_2) \, .
\ee
Making use of the fact that in the deflationary phase the Hubble crossing condition
gives
\be
\eta(-t_3(k)) \, \sim \, k^{-1}
\ee
we immediately obtain
\be \label{result}
\zeta(k, -t_3(k)) \, \sim \, k^{-7/2} \, .
\ee
Our result (\ref{result}) shows that the initial vacuum spectrum has been
transformed into a extremely red spectrum. The redness of the spectrum is
a consequence of the fact that $\zeta$ is growing on super-Hubble scales
in the contracting phase, and that long wavelength modes are super-Hubble
for a much longer time than short wavelength modes.

\subsection{Fluctuations during and after the Bounce}

During the bounce, the variable $v_k$ is oscillating, hence $\zeta$ will
have the same amplitude at Hubble radius crossing during inflation
after the bounce as it had at Hubble radius crossing before the bounce.
After Hubble radius crossing in the expanding phase, $\zeta$ is
conserved. Hence at late tiimes $t$
\be
\zeta(k, t) \, = \, \zeta(k, t_3(k)) \, \simeq \, \zeta(k, -t_3(k)) \, \sim \, k^{-7/2} \, ,
\ee
which corresponds to a power spectrum
\be
P_{\zeta}(k) \, \equiv \, \frac{k^3}{12 \pi^2} |\zeta(k)|^2 \, \sim \, k^{-4} \, ,
\ee
i.e. spectral index $n_s = -3$.
This is a manifestation of the severity of the trans-Planckian problem for
fluctuations in inflationary cosmology. Without new physics affecting
the evolution of fluctuations, we have a model in which fluctuations
emerging from inflation are highly non-scale-invariant.

\section{Back-Reaction Effects}

As emphasized in \cite{Tanaka,Starob} and explored in more
detail in \cite{Jerome3}, the magnitude of trans-Planckian effects
which grow towards the UV end of the spectrum is bounded by
its back-reaction on the background. In the following we will
see that back-reaction effects tightly constrain the length of
a deflationary contracting phase if the modes which are
entering the Hubble radius during the deflationary period
are not in their (Bunch-Davies) vacuum state.

We can estimate the energy density $\rho_f$ in sub-Hubble scale
perturbations in the following way.  
First of all, recall (see e.g. \cite{MFB,RHBrev1}) that for scalar field matter
\be
v \, = \, a \bigl( \delta \varphi + \frac{z}{a} \Phi \bigr) \, .
\ee
Since on sub-Hubble scales the energy density is dominated by matter
and since the matter fields are oscillating, the energy density can be well
approximated by the gradient energy of $\delta \varphi$. Since on
sub-Hubble scales we can approximate $\delta \varphi = a^{-1} v$,
and making use of the fact that $v_k$ oscillates with constant
amplitude, we have
\be
\rho_f(t) \, \simeq \, a^{-2}(t) 
\int_{k_{\rm{min}}(t)}^{k_{\rm{max}}} d^3{\bf{k}} a^{-2}(t) k^2 v(k, \eta_3(k))^2 \, ,
\ee
where $k_{\rm{max}}$ corresponds to the wavelength which crosses
the Hubble radius at time $-t_2$ and $k_{\rm{min}}$ corresponds
to the mode which is entering the Hubble radius at time $t$.

Making use of the scaling of $v$ in the radiation phase and in the
deflationary phase we find
\be
v(k, \eta_3(k)) \, = \, v(k, \eta_2) \bigl( \frac{\eta_3(k)}{\eta_2} \bigr)^2
\, = \, v(k, \eta_1(k))  \bigl( \frac{\eta_3(k)}{\eta_2} \bigr)^2
\ee
and hence
\bea
\rho_f(t) \, &\simeq& \, a^{-4}(t) \int_{k_{\rm{min}}(t)}^{k_{\rm{max}}} d^3{\bf{k}} k^2
|v(k, \eta_1(k)) |^2 \bigl( \frac{\eta_3(k)}{\eta_2} \bigr)^4  \nonumber \\
&\sim& \,  a^{-4}(t) \int_{k_{\rm{min}}(t)}^{k_{\rm{max}}} dk k^4 k^{-1} \bigl( \frac{k_{\rm{max}}}{k} \bigr)^4
\nonumber \\
&\sim& \, a^{-4}(t) k_{\rm{max}}^4 \rm{ln}(\frac{k_{\rm{max}}}{k_{\rm{min}}})\, . 
\eea
Since $a^{-1}(t_2) k_{\rm{max}} = H$, we find
\be
\rho_f(t) \, \simeq \,  \bigl( \frac{a(t_2)}{a(t)} \bigr)^4 H^4 \rm{ln}(\frac{k_{\rm{max}}}{k_{\rm{min}}})\, ,
\ee
which is smaller than the background energy density provided that the period
of deflation is short. 

As expected, the energy density of these sub-Hubble modes
scales as radiation. Hence, as soon as
\be
 \bigl( \frac{a(t_2)}{a(t)} \bigr) \, > \, \bigl( \frac{m_{pl}}{H} \bigr)^{1/2}
 \ee
 (up to logarithmic factors) then the radiation in the fluctuation modes will begin to
 dominate over the constant energy density driving deflation, and a
 transition to another radiative contraction phase will set in. Hence, the
 number $\tilde{N}$ of e-foldings of the deflationary phase is bounded by
 \be
 \tilde{N} \, < \, \frac{1}{2} \rm{log} (\frac{m_{pl}}{H}) \, ,
 \ee
 where $m_{pl}$ denotes the Planck mass. Thus, if we want to have
 a sufficiently long period of inflation in the expanding phase, enough
 for inflation to solve the problems of Standard Big Bang cosmology
 such as the horizon and flatness problems \cite{Guth}, then we
 cannot have a symmetric bounce.

Another consequence of this result is that the range of wavelengths
which exhibit the $n = -3$ spectrum is bounded. If $N$ denotes
the number of e-foldings of inflation in the expanding phase,
then only modes which exit the Hubble radius during the
inflationary phase more than $N - {\tilde N}$ e-foldings before
the end of inflation exhibit the modified spectrum. Modes which
exit the Hubble radius later are in the usual Bunch-Davies since
they were never super-Hubble during the contracting phase.

We conclude that if the inflationary phase lasted a large number
of e-foldings, then the trans-Planckian effects from the
contracting phase are red-shifted to wavelengths which are
still super-Hubble today. This result is analogous to that
obtained in \cite{Cline} who show that initial condition
signatures which carry deviations from the usual scale-invariance
of the spectrum of cosmological perturbations are red-shifted
to the far infrared as $N$ increases. 

However, in the context of a bouncing cosmology, we are not 
free to simply choose initial conditions for the scalar field $\varphi$
driving inflation such that a very large number of e-foldings
results. If we start, as assumed here, with the scalar field
oscillating in the far past in the contracting phase, then it is
very unlikely that a large value of $\varphi$ will be generated
after the bounce. Hence, we expect that $N$ will be close to
the minimal value required for inflation to solve the problems
of Standard Big Bang cosmology, and hence the trans-Planckian
signatures discussed in this paper will be on observable
scales

\section{Conclusions and Discussion}

We have presented a model which demonstrates the severity of the
trans-Planckian problem for cosmological fluctuations in inflationary
cosmology. Embedding a period of inflationary expansion into
a non-singular model with a cosmological bounce which contains
a deflationary phase of contraction,
we have shown that vacuum initial conditions for fluctuations in
the early stages of the contracting universe leads to a spectrum of
perturbations which emerge from the period of inflationary expansion
which is not scale-invariant. In the example in which the initial
phase of contraction is dominated by radiation, we have shown that
a power spectrum with spectral index $n_s = -3$ results.

Our analysis does not make use of any non-standard evolution of the
fluctuations - these are traced from the time when they first exit the
Hubble radius in the contracting phase to the present time using
the general relativistic fluctuation equations.

By the nature of the trans-Planckian problem for cosmological
fluctuations, we cannot trust the validity of the fluctuation
equations on length scales smaller than the Hubble scale. What
we have shown, however, is that since the fluctuations dip into
the trans-Planckian region during the contracting deflationary
phase with a spectrum which is very different from the
vacuum spectrum, it would require un-natural fine-tuning of
trans-Planckian physics to convert the fluctuations into a
spectrum which looks vacuum-like during the expanding
inflationary phase.

We have seen that back-reaction effects of fluctuations tightly
constrain the duration of the deflationary phase before the
bounce. This constrains the range of fluctuation modes
which are subject to the trans-Planckian corrections
discussed here. If the deflationary phase lasts $\tilde{N}$
e-foldings, and the inflationary phase of expansion $N$
e-foldings, then only modes which exit the Hubble radius
during the expanding period more than $N - \tilde{N}$
e-foldings before the end of inflation exhibit the
modified spectrum. However, in the context of our setup, it 
is highly improbable to get a long period of inflation. Hence, it
is very likely that the trans-Planckian signatures will be
in the observable range of wavelengths.

In conclusion, we have presented an inflationary model with a 
preceding cosmological bounce in which cosmological fluctuations 
emerge with a $n_s = -3$ spectrum
and not with a scale-invariant spectrum. This demonstrates that inflationary
cosmology must deal with the trans-Planckian problem for fluctuations. Conversely,
our results also imply that if our universe in fact underwent a period of inflation,
then Planck-scale physics can be probed with current cosmological observations.

\begin{acknowledgments}

We wish to thank Yi-Fu Cai and Tao-Tao Qiu for discussions. RB wishes to thank 
Matt Kleban for stimulating questions on the issue of
back-reaction. He also thanks the
Theory Division of the Institute of High Energy Physics (IHEP) for
their wonderful hospitality and financial support, and the KITPC for
hospitality during the finalization of this draft. RB is 
supported by an NSERC Discovery Grant and by the Canada Research
Chairs Program. The research of X.Z. is supported  in part by the
National Science Foundation of China under Grants No. 10533010,
10675136 and 10821063, by the 973 program No. 2007CB815401, and 
by the Chinese Academy of Sciences under Grant No. KJCX3-SYW-N2

\end{acknowledgments}

\end{document}